\documentclass[copyright,creativecommons]{eptcs}
\providecommand{\event}{PLACES 2009} 
\providecommand{\volume}{??}
\providecommand{\anno}{2009}
\providecommand{\eid}{??}

\usepackage[utf8]{inputenc}
\usepackage{listings}
\usepackage{xspace}
\usepackage{graphics, graphicx}
\usepackage{amsmath,amsthm}
\usepackage{stmaryrd}

\usepackage{listings}

\def\eg{\emph{e.g.}\;}
\def\ie{\emph{i.e.}\;}
\def\cf{\emph{c.f.}\;}

\newcommand{\keyw}[1]{\textbf{\textsf{#1}}\,}
\newcommand{\inaction}{\textbf{\textsf{0}}}

\newcommand{\installk}{\keyw{install}}

\newcommand{\ink}{\keyw{in}}
\newcommand{\systemk}{\keyw{extern}}

\newcommand{\letk}{\keyw{let}}

\newcommand{\sendk}{\keyw{send}}

\newcommand{\firek}{\keyw{timer}}
\newcommand{\everyk}{\keyw{every}}
\newcommand{\expirek}{\keyw{expire}}
\newcommand{\receivek}{\keyw{receive}}

\newcommand{\branek}{\keyw{sensor}}

\newcommand{\selfk}{\text{self}}

\newcommand{\sensor}[6]{[{#1} \triangleright {#2}]^{{#3}, {#4}}_{{#5}, {#6}}}
\newcommand{\sensord}{\sensor {P,R} {M, T} I O t p}
\newcommand{\tagsensor}[7]{[{#1}\triangleright{#2}]^{{#3}, {#4}}_{{#5}, {#6}}\{{#7}\}}
\newcommand{\tagsensord}{\tagsensor {P,R} {M,T} I O p t S}

\newcommand{\amoduled}{\{ {l_i} = {\abstr {\vec x_i}{P_i}}\}_{i\in I}}
\newcommand{\abstr}[2]{({#1})\,{#2}}

\newcommand{\invk}[3]{{#1}.{#2}({#3})}
\newcommand{\invkd}{\invk {v}{l}{\vec v}}
\newcommand{\system}[1]{\systemk\ {#1}}
\newcommand{\systemd}{\system {\msgd}}
\newcommand{\fire}[3]{\firek\ {#1}\ \everyk\ {#2}\ \expirek\ {#3}}

\newcommand{\fired}{\fire {\msgd} v v}

\newcommand{\send}[1]{\sendk\ {#1}}
\newcommand{\sendd}{\send {\msgd}}
\newcommand{\msg}[2]{{#1}({#2})}
\newcommand{\msgd}{\msg l {\vec v}}
\newcommand{\receivedd}{\receivek}

\newcommand{\install}[2]{{#1}.\installk\ {#2}}
\newcommand{\installd}{\install v v}
\newcommand{\Let}[3]{\letk\ {#1} = {#2}\ \ink\ {#3}}
\newcommand{\Letd}{\Let x P P}

\newcommand{\parn}{\,\vert\,}

\newcommand{\ldsq}{[\![} 
\newcommand{\rdsq}{]\!]} 
\def\context#1{\mathcal{C}\ldsq#1\rdsq}
\def\emptycontext{[\;]}

\def\obj#1{\{#1\}}
\def\typeamodd{\obj {l_i \colon \vec \tau_i \rightarrow \tau_i}_{i \in I}}
\def\typesmodd{\anyObj {l_i \colon \vec \tau_i \rightarrow \tau_i}_{i \in I}}

\def\senObj#1{[#1]}

\def\anyObj#1{\langle {#1} \rangle}

\newcommand{\functionType}[2]{{#1} \rightarrow {#2}}
\newcommand{\functionTyped}{\functionType {\vec \tau} {\tau}}
\newcommand{\recType}[2]{\mu {#1}.{#2}}
\newcommand{\recTyped}{\recType{\alpha} {\tau}}
\newcommand{\dom}{\operatorname{dom}}
\newcommand{\disj}{\operatorname{,}}

\newcommand{\dist}{\operatorname{inRange}}
\newcommand{\route}{\operatorname{networkRoute}}

\newcommand{\fv}{\operatorname{fv}}

\newcommand{\methJoin}{+}
\newcommand{\subs}[2]{[{#1}/{#2}]}

\newcommand{\congr}{\equiv}
\newcommand{\reduces}{\rightarrow}

\newcommand{\reducesn}{\reduces^{*}}

\newcommand{\err}{\overset{\text{err}}{\longmapsto}}

\newcommand{\mkRrule}[1]{{\footnotesize \textsc{R-#1}}}
\newcommand{\mkTrule}[1]{{\footnotesize \textsc{T-#1}}}
\newcommand{\mkSrule}[1]{{\footnotesize \textsc{S-#1}}}
\newcommand{\mkErule}[1]{{\footnotesize \textsc{E-#1}}}
\newcommand{\Rfunction}{\mkRrule{function}}

\newcommand{\RinstallS}{\mkRrule{install-sensor}}
\newcommand{\RinstallM}{\mkRrule{install-module}}
\newcommand{\Rlet}{\mkRrule{let}}

\newcommand{\Rfire}{\mkRrule{timer}}
\newcommand{\RcallS}{\mkRrule{call-sensor}}
\newcommand{\RcallM}{\mkRrule{call-module}}
\newcommand{\Rnocall}{\mkRrule{no-function}}
\newcommand{\Rtrigger}{\mkRrule{trigger}}
\newcommand{\Rdiscard}{\mkRrule{expire}}
\newcommand{\Rsend}{\mkRrule{send}}
\newcommand{\Rreceive}{\mkRrule{receive}}
\newcommand{\Rsystem}{\mkRrule{extern}}
\newcommand{\Rnext}{\mkRrule{next}}
\newcommand{\Ridle}{\mkRrule{idle}}
\newcommand{\RreceiveN}{\mkRrule{no-message}}

\newcommand{\Rmove}{\mkRrule{move}}

\newcommand{\Rnetwork}{\mkRrule{network}}

\newcommand{\Rcongr}{\mkRrule{congr}}
\newcommand{\Rbroadcast}{\mkRrule{broadcast}}
\newcommand{\Rrelease}{\mkRrule{release}}

\newcommand{\SmonoidSensor}{\mkSrule{monoid-Sensor}}
\newcommand{\SinitSend}{\mkSrule{init-Send}}

\newcommand{\type}{\vdash}
\newcommand{\is}{\colon}
\newcommand{\TSinaction}{\mkTrule{off}}
\newcommand{\Tparallel}{\mkTrule{par}}

\newcommand{\TrunQueue}{\mkTrule{run-queue}}

\newcommand{\TcommQueue}{\mkTrule{comm-queue}}
\newcommand{\TeventQueue}{\mkTrule{event-queue}}

\newcommand{\TSsensor}{\mkTrule{sensor}}
\newcommand{\TSbSensor}{\mkTrule{bSensor}}

\newcommand{\Tlet}{\mkTrule{let}}

\newcommand{\Tfire}{\mkTrule{timer}}
\newcommand{\Treceive}{\mkTrule{receive}}
\newcommand{\Tcall}{\mkTrule{call}}
\newcommand{\TsCall}{\mkTrule{extern}}

\newcommand{\TinstS}{\mkTrule{sInstall}}
\newcommand{\TinstM}{\mkTrule{mInstall}}
\newcommand{\Tbcast}{\mkTrule{send}}
\newcommand{\TVvar}{\mkTrule{var}}
\newcommand{\TVlabel}{\mkTrule{label}}
\newcommand{\TVseq}{\mkTrule{seq}}
\newcommand{\TVsensor}{\mkTrule{sensor}}

\newcommand{\Tcode}{\mkTrule{code}}

\newcommand{\Tbin}{\mkTrule{built-in}}

\newcommand{\seqSets}[3]{{#1}; {#2}; {#3}}
\newcommand{\seqSetsd}{\seqSets {\tau_S} {\tau_M} {\Gamma}}
\newcommand{\arrowType}[2]{{#1} \rightarrow {#2}}
\newcommand{\arrowTyped}{\arrowType {\vec \tau} {\tau}}

\newcommand{\Efunction}{\mkErule{cFunction}}
\newcommand{\Ecall}{\mkErule{call}}
\newcommand{\Einstall}{\mkErule{install}}
\newcommand{\Epar}{\mkErule{par}}
\newcommand{\Estr}{\mkErule{str}}
\newcommand{\nerr}{\overset{\text{err}}{\,\,\arrownot\!\!\longmapsto}}

\newcommand{\pad}{\;\;}
\newcommand{\Space}[1]{\pad{#1}\pad}
\newcommand{\grmeq}{\Space{::=}}

\newcommand{\grmor}{\;\mid\;}

\newcommand{\myparagraph}[1]{\vspace*{0.2cm}\noindent\textbf{#1}}

\newtheorem{definition}{Definition}

\newtheorem{theorem}{Theorem}
\newtheorem{lemma}[theorem]{Lemma}

\newtheorem{corollary}[theorem]{Corollary}

\newcommand{\rulespace}{0.2cm}

\newcommand{\noEvent}[2]{\mathop{\mathrm{noEvent}}\nolimits ({#1}, {#2})}
\newcommand{\noEventd}{\noEvent T t}

\def\authorrunning{Martins, Lopes, and Barros}
\def\titlerunning{Towards Safe Programming of WSN}

\begin{document}

\title{Towards the Safe Programming of \\ Wireless Sensor Networks}

\author{Francisco Martins
\institute{LASIGE \& DI-FCUL,\\
   Lisbon, Portugal
   \email{fmartins@di.fc.ul.pt} 
}
\and
Lu\'{\i}s Lopes
\institute{CRACS/INESC-Porto \& DCC-FCUP,\\
  Porto, Portugal
  \email{lblopes@dcc.fc.up.pt}
}
\and
Jo\~ao Barros
\institute{IT \& FEUP,\\
  Porto, Portugal
  \email{jbarros@fe.up.pt}
}
}
\maketitle
\begin{abstract}
Sensor networks are rather challenging to deploy, program, and
debug. Current programming languages for these platforms suffer from a
significant semantic gap between their specifications and underlying
implementations. This fact precludes the development of (type-)safe
applications, which would potentially simplify the task of programming and
debugging deployed networks. In this paper we define a core calculus
for programming sensor networks and propose to use it as an assembly
language for developing type-safe, high-level programming languages.
\end{abstract}

\textbf{keywords}: Sensor Networks, Programming Languages,
Process-Calculi.


\section{Introduction and Motivation}
\label{sec:introduction}

Wireless sensor networks are composed of huge numbers of small
physical devices capable of sensing the environment and connected
using ad-hoc networking protocols over radio links~\cite{survey:akyildiz:etal:02}.
These platforms have several unique characteristics when compared with
other ad-hoc networks.
First, sensor networks are often designed for specific applications or
application domains making software re-usability and portability an
issue.
Sensor devices have very limited processing power (CPU), available
memory, and battery lifetime, and are often deployed at remote
locations making physical access to the devices (\eg for maintenance)
difficult or even impossible.
%
%
For these reasons, programming such large scale distributed systems
can be daunting. Programs must be \emph{lightweight}, produce a
\emph{small memory footprint}, be \emph{power conservative}, be
\emph{self-reconfigurable} (\ie may be reprogrammed dynamically
without physical intervention on the devices) and, we argue, be
\emph{(type-)safe}.

To date several programming languages and run-time systems have been
proposed for wireless sensor networks (see~\cite{bookchapter} and
references therein) that address some of the above issues, but few
tackle the \emph{safety} issue. Regiment~\cite{regiment2}, a strongly
typed functional \emph{macroprogramming} language, is the closest to
achieve this goal by providing a type-safe compiler.
However, Regiment is then compiled into a low-level \emph{token
 machine language} that is not type-safe. This intermediate language
is itself compiled into a nesC implementation of the run-time based on
the \emph{distributed token machine} model, for which no safety
properties are available.
In fact, in general, an underlying model with well-studied operational
semantics for sensor networks seems to be lacking. The absence of such
a model reveals itself as a considerable semantic gap between the
semantics of the (sometimes high-level) programming languages and their
respective implementations.

In this paper we propose Callas, a calculus for programming sensor
networks, based on the formalism of process
calculi~\cite{async-pi:honda:tokoro:91,pi:milner:parrow:walker:92},
that aims to establish a basic computational model for sensor
networks. The goal is to diminish the above mentioned semantic gap
by proceeding bottom-up, using Callas as a basic assembly language
upon which high-level programming abstractions may be encoded as
semantics preserving, derived constructs.
Callas is an evolution from a previous proposal~\cite{sensorcomm07} by
the authors, which unlike its original sibling  provides: (a)
decoupled semantics for in-sensor computation (associated with the
\emph{application layer}) and networking (associated with the
\emph{data-link} and \emph{network} layers); b) support for a form of
timed \emph{events}; and c) \emph{event-driven} semantics.



\section{Overview of Callas}
\label{sec:calculus}

The syntax of Callas is provided by the grammar in
Figure~\ref{fig:syntax}. Let $\vec \alpha$ denote a possibly empty
sequence $\alpha_1 \dots \alpha_n$ of elements of some syntactic
category $\alpha$. We let~$l$ range over a countable set of
\emph{labels} representing function names, and let $x$ range
over a countable set of \emph{variables}. These sets are pairwise
disjoint.


\begin{figure}
  \begin{equation*}
    \begin{aligned}
      S & \grmeq                          & & \text{\emph{Sensors}}\\
        & \quad \; \inaction           & & \text{empty network}\\
        & \grmor S \parn S                & & \text{composition}\\
        & \grmor \sensor {P,R} {M,T} I O p t  & & \text{sensor}\\[0.5cm]
      v & \grmeq                          & & \text{\emph{Values}}\\
        & \quad \; b                   & & \text{built-in value}\\
        & \grmor x                        & & \text{variable}\\
        & \grmor M                        & & \text{module}\\
        & \grmor \branek                  & & \text{installed functions}\\[0.4cm]
      m & \grmeq \langle l(\vec v)\rangle     & & \text{messages}\\[0.4cm]
      R & \grmeq P_1::\dots::P_n  & & \text{run-queue} \\
      I,O & \grmeq m_1::\dots::m_n        & & \text{message queues} 
    \end{aligned}
    \qquad
    \begin{aligned}
      M & \grmeq              & & \text{\emph{Modules}}\\                
        & \quad \; \amoduled  & & \text{module}\\[0.5cm]
      P & \grmeq              & & \text{\emph{Processes}}\\
        & \quad \; v       & & \text{value}\\
        & \grmor \invkd       & & \text{function call}\\
        & \grmor \systemd     & & \text{external call}\\
        & \grmor \fired       & & \text{timed call}\\
        & \grmor \sendd       & & \text{communication}\\
        & \grmor \receivedd   & & \text{communication}\\
        & \grmor \installd    & & \text{install code}\\
        & \grmor \Letd        & & \text{sequence}\\[0.3cm]
      T & \grmeq \{(l_i(\vec v_i),v_i,v_i,v_i)\}_{i \in I} & & \text{timed calls} 
    \end{aligned}
  \end{equation*}
\caption{The syntax of Callas.}
\label{fig:syntax}
\end{figure}


A network $S$ is an abstraction for a network of real-world sensors
connected via radio links. We write it as a flat, unstructured
collection of sensors combined using the parallel composition
operator. The empty network is represented by symbol $\inaction$.
A sensor $\sensord$ is an abstraction for a sensor device.  It
features a running process~($P$) and a double-ended
queue of processes scheduled for execution~($R$). Its memory stores
both the installed code for the application~($M$) and a table of
timers for function calls~($T$). These components represent the
application layer of the protocol stack for the sensor. The interface
with the lower level networking and data-link layers is modeled using
incoming~($I$) and outgoing~($O$) queues of messages.
The sensors have a measurable position~($p$) and their own
clocks~($t$), and are able to measure some physical property
(e.g.\@ temperature, humidity) by calling appropriate \emph{external}
functions.
The code in $M$ consists of a set of named functions.
The syntax $l = (\vec x) P$ represents a function, where $l$ is the name,
$(\vec x)$ the parameters, and $P$ the body.
The $I$ ($O$) queue buffers messages received from (sent to) the
network. Messages are just packaged function calls
$\langle l(\vec v)\rangle$.
Finally, $T$ is a set that keeps information on timers for function
calls. For each timer, a tuple is maintained with the following
information: the call to be triggered, the timer period, the
time after which the timer expires and, the time of the next
call.

A process $P$ can be one of the following: 
(a) a value $v$ that represents
the data exchanged between sensors. It can be a basic value ($b$) that
can intuitively be seen as the primitive data types supported by the
sensor's hardware or a module ($M$). The special value $\branek$
represents the module that holds the functions installed at the
sensor; 
(b) a synchronous call $\invkd$ to a function $l$ in a module $v$;
(c) a synchronous external call -- $\systemd$; 
(d) a timer -- $\fired$, that
calls an installed function $l(\vec v)$ periodically, controlled by a
timer; 
(e) an asynchronous remote call -- $\sendd$, that adds a message
$\langle l(\vec v)\rangle$ to the outgoing queue ($O$); 
(f) a receptor --
$\receivedd$, that gets a message from the incoming queue ($I$); 
(g) a module installation -- $\install v{v'}$, that adds the set of
functions in $v'$ to $v$; and, finally 
(h) a $\letk$ construct that
allows the processing of intermediate values in computations.
The latter is also useful to derive a basic sequential composition
construct (in fact, $\letk\ x = P\ \ink\ P' \equiv P~;~P'$ with $x
\not\in \fv(P')$). We make frequent use of this construct to impose a
more imperative style of programming.
%
%
Each function in a module has as the first parameter the variable
\selfk{} that is, as usual, a reference to the current module, i.e.,
the one the function belongs to. Each call to a function $v.l(\vec v)$
passes $v$ as the first argument in $l(\vec v)$.

In the sequel we present two small examples of programs written in
Callas. Both examples have two components: the code to be run at a
base-station (\emph{sink}) and the code to be run at each of the
other nodes (\emph{sensor}).

\textbf{Streaming data.} The program that runs on the \emph{sink}
starts by installing, in the local memory ($M$), a module with a
\lstinline{receiver} function and a \lstinline{gather} function. The
former just listens for messages from the network on the incoming
queue. The latter simply logs the arguments using a built-in external
call. Then, it starts a timer for the \lstinline{receiver} function
with a period of 5 milliseconds for 10 seconds. Finally, the sink
broadcasts a \lstinline{setup} message with a period of 100
milliseconds and a duration of 10 seconds. The call is placed in the
outgoing queue of the sink ($O$). In these examples we write
\lstinline{install} as a compact form for \lstinline{sensor.install}.\\

\begin{lstlisting}
// sink
install { 
   receiver = (self) 
      receive
   gather = (self,x,y) 
      external log(x,y)             };
timer receiver() every 5 expire 10000;
send setup(100,10000)

// sensor
install { 
   receiver = (self) 
      receive
   setup  = (self,x,y) 
      timer sample() every x expire y
   sample = (self) 
      let x = external time() in 
      let y = external data() in 
      send gather(x,y)              };
timer receiver() every 5 expire 10000;
\end{lstlisting}

Each \emph{sensor} starts by installing a module with a
\lstinline{receiver} function, similar to that on the sink, and
\lstinline{setup} and \lstinline{sample} functions. Then it starts a
timer on \lstinline{receiver} and waits for incoming messages. When a
sensor receives a \lstinline{setup} message from the network, it sets
up another timer to periodically call \lstinline{sample} in the same
module. When this function is executed the local time and the desired
data are read with external calls and a \lstinline{gather} message is
sent to the network carrying those values.

Note that the routing of messages is transparent at this level. It is
controlled at the network and data-link layers and we model this by
having an extra semantic layer for the network
(\cf Figure~\ref{fig:reduction-net}). In this example, the messages
from the sink are delivered to every sensor that carries a
\lstinline{setup} function. The information originating in the
sensors, in the form of \lstinline{gather} messages, on the other
hand, is successively relayed up to the sink (since sensors have no
\lstinline{gather} functions implemented).

\textbf{The maximum value of a data attribute and the MAC address of
 the sensor that reads it.} This example follows much the same
principles of the above, except that it is a single shot
request. Instead of computing the maximum value of the data attribute
only at the sink, we optimize the program so that each sensor has two
attributes \lstinline{max_data} and \lstinline{max_mac} that keep,
respectively, the maximum value for the data that passed through the
sensor, and the associated MAC address.\\

\begin{lstlisting}
// sink
install { 
   receiver = (self) 
      receive
   gather   = (self,x,y) 
      external log(x,y)             };
timer receiver() every 5 expire 10000;
send setup()

// sensors
install { 
   receiver = (self) 
      receive
   setup = (self)     
      let x = external data() in 
      let y = external  mac()  in 
      self.install { max_data = (self) x 
                     max_mac  = (self) y };
      send gather (x,y)
   gather = (self,x,y) 
      let val = self.max_data() in 
      if x > val then 
         self.install { max_data = (self) x 
                        max_mac  = (self) y };
         send gather(x,y);          };
timer receiver() every 5 expire 10000;
\end{lstlisting}

The program that runs on the \emph{sink} is very similar to that of
the previous example. After installing the \lstinline{receiver} and
the \lstinline{gather} functions, it starts the receiver and
broadcasts a \lstinline{setup} message to the network.

The \emph{sensors} get the call from the network using their receivers
and execute \lstinline{setup}. The data and MAC address are obtained
by calling external functions and sent to the network in
\lstinline{gather} messages. Each time such a message is relayed by a
sensor on its way to the sink, the relaying sensor checks whether it
is worth to send the data forward by comparing it with the local
maximum. This strategy manages to substantially reduce the required
bandwidth at the sensors closest to the sink. The sink implementation
of \lstinline{gather} stops the relaying and logs the data. Note that
in this example, to simplify, more than one maximum value may be
recorded at the sink. Also, we use an \lstinline{if-then} construct
that is not provided in the base calculus but that can easily be added
for convenience.

Unlike the previous example, here every sensor will relay
\lstinline{gather} messages only after some internal processing, by
its own version of the homonym function.

\myparagraph{Semantics.}
\label{sec:semantics}
%
The calculus has two variable binders: the $\letk$ and the function
constructs, inducing the usual definition for free and bound variables.
The displayed occurrence of variable $x$ is a \emph{binding} with
\emph{scope}~$P$ both in $\letk\ x=P'\ \ink\ P$ and in
$l = (\dots, x, \dots) P$.
An occurrence of a variable is \emph{free} if it is not in the scope of a
binding.
Otherwise, the occurrence of the variable is \emph{bound}.
The set of free variables of a sensor $S$ is referred to as $\fv(S)$.

We present the reduction relation with the help of a structural
congruence, as it is usual~\cite{ccs:milner:80},
given in Figure~\ref{fig:congruence}.
Here, $\SinitSend$ is the only non-standard rule and provides a sensor
with a conceptual \emph{membrane} that engulfs neighboring sensors as
they become engaged in communication. This prevents the reception of
duplicate copies of the \emph{same} message from the source sensor
during a transmission.
%

\begin{figure}
  \begin{gather*}
    \tag{\SmonoidSensor} S_1 \parn S_2 \congr S_2 \parn S_1, \qquad S \parn \inaction \congr S, \qquad  S_1 \parn (S_2 \parn S_3)  \congr (S_1 \parn S_2) \parn S_3 \\ \\
    \sensor {P,R} {M,T} I O p t \congr \tagsensor {P,R} {M,T} I O p
t {\inaction} \qquad \tag{\SinitSend}
  \end{gather*}
\caption{Structural congruence for sensors.}
\label{fig:congruence}
\end{figure}

%
The reduction relation is inductively defined by the rules in
Figures~\ref{fig:reduction} and~\ref{fig:reduction-net}.
Since processes evaluate to values, we allow for reduction within the
$\letk$ construct and therefore present the reduction relation using
the following reduction contexts: $\context{\cdot} \grmeq \emptycontext
\grmor \letk\ x=\context{\cdot} \;\ink\ P$.
The reduction in a sensor is driven by running process~$P$.


\begin{figure}
  \begin{gather*}
    \tag{\Rsystem}
    \frac{
      \noEventd
    }{
      \sensor {\context{\systemd},R} {M,T} I O p t 
      \reduces
      \sensor {\context{v},R} {M, T} I O p {t+1} 
    }
    \\[\rulespace]
    \tag{\RinstallS}
    \frac{
      \noEventd
    }
    {
      \sensor {\context{\install {\branek} {M'}},R} {M, T} I O p t 
      \reduces
      \sensor {\context{\obj{}},R} {M+M', T} I O  p {t+1} 
    }
    \\[\rulespace]
    \tag{\RinstallM}
    \frac{
      \noEventd
    }
    {
      \sensor {\context{\install {M'} {M''}},R} {M, T} I O p t 
      \reduces
      \sensor {\context{M' + M''},R} {M, T} I O  p {t+1} 
    }
    \\[\rulespace]
    \tag{\Rsend}
    \frac{
      \noEventd
    }
    {
      \sensor {\context{\sendd},R} {M, T} I O p t 
      \reduces
      \sensor {\context{\obj{}},R} {M, T} I {O::\langle \msgd \rangle} p {t+1}
    }
    \\[\rulespace]
    \tag{\Rreceive}
    \frac{
       \noEventd
    }
    {
      \sensor {\context{\receivedd},R} {M, T} {\langle \msgd \rangle :: I} O p t 
      \reduces
      \sensor {\context{\obj{}},R:: \invk \branek l {\vec v}} {M, T} I O p {t+1}
    }
   \\[\rulespace]
    \tag{\RreceiveN,\Ridle}
    \frac{
       \noEventd
    }
    {
      \sensor {\context{\receivedd},R} {M, T} {\varepsilon} O p t 
      \reduces
      \sensor {\context{\obj{}},R} {M, T} I O p {t+1}
    }
    \qquad
    \frac{
      \noEventd
    }
    {
      \sensor {v,\varepsilon} {M, T} I O p t 
      \reduces
      \sensor {v,\varepsilon} {M, T} I O p {t+1}
    }
   \\[\rulespace]
    \tag{\Rnext,\Rmove}
    \frac{
      \noEventd
    }
    {
      \sensor {v,P::R} {M, T} I O p t 
      \reduces
      \sensor {P,R} {M, T} I O p {t+1}
    }
    \qquad
    \frac{
      \noEventd
    }
    {
      \sensor {P,R} {M, T} I O p t 
      \reduces
      \sensor {P,R} {M, T} I O {p'} t
    }
    \\[\rulespace]
    \tag{\Rlet}
    \frac{
      \noEventd
    }
    {
      \sensor {\context{\Let x v P},R} {M, T} I O p t
      \reduces
      \sensor {\context{P\subs v x},R} {M, T} I O p t 
    }
    \\[\rulespace]
    \tag{\RcallS}
     \frac{
      M(l) = (\selfk\ \vec x) P
       \quad 
       \noEventd
     }{
       \sensor {\context{\invk \branek l {\vec v}},R} {M,T} I O p t 
       \reduces
       \sensor {\context{P\subs {M\ \vec v} {\selfk\ \vec x}},R} {M, T} I O p {t+1} 
     }
     \\[\rulespace]
    \tag{\RcallM}
     \frac{
       M'(l) = (\selfk\ \vec x) P \quad  \noEventd
     }{
       \sensor {\context{\invk {M'} l {\vec v}},R} {M,T} I O p t 
       \reduces
       \sensor {\context{P\subs {M'\ \vec v} {\selfk\ \vec x}},R} {M, T} I O p {t+1} 
     }
     \\[\rulespace]
     \tag{\Rnocall}
     \frac{
       l \not\in \dom(M) \quad  \noEventd
     }{
       \sensor {\context{\invk \branek l {\vec v}},R} {M,T} I O p t 
       \reduces
       \sensor {\obj{},R::\context{\invk \branek l {\vec v}}} {M, T} I O p {t+1} 
     }
    \\[\rulespace]
    \tag{\Rfire}
     \frac{
       T' = T \uplus (\msgd, v, t + v', t + v)
       \quad \noEventd
     }{
       \sensor {\context{\fire \msgd {v} {v'}},R} {M,T} I O p t 
       \reduces
       \sensor {\context{\obj{}},\invk \branek l {\vec v}::R} {M,T'} I O p {t+1} 
     }
     \\[\rulespace]
    \tag{\Rtrigger}
     \frac{
       t \le v'  \quad  T' = T \uplus (\msgd, v, v', t + v) 
     }{
       \sensor {P,R} {M,T \uplus (\msgd, v, v', t)} I O p t 
       \reduces
       \sensor {P,\invk \branek l {\vec v}::R} {M, T'} I O p t 
     }
     \\[\rulespace]
    \tag{\Rdiscard}
     \frac{
       t > v'
     }{
       \sensor {P,R} {M,T \uplus (\msgd, v, v', t)} I O p t 
       \reduces
       \sensor {P,R} {M, T} I O p t 
     }
\end{gather*}
\center See Definition~\ref{def:plus} for the formal meaning of operator $+$.
\caption{Reduction semantics for sensors.}
\label{fig:reduction}
\end{figure}


Within sensors reduction proceeds without obstacle while the internal
clock $t$ is not such that a timed call must be triggered. This is
controlled by the predicate $\underline{\mathop{\mathrm{noEvent}}}$
that checks the time of the next activation for every timed call
against the current time.
There is no special reason
why the increments in the clock are unitary. One could easily assume
that each instruction consumes a different number of processor cycles
and reflect that scenario in the rules. Some rules (\eg \Rlet{})
simply re-structure a process and thus we assume that no cycles are
consumed.

Rule \Rsystem{} calls a synchronous external function and receives a
value as the result. The rules \RinstallS{} and \RinstallM{} handle
module updates. The former takes the module with the code installed at
the sensor and updates it with the code of another module $M'$. The
resulting new module is installed in the sensor. The latter applies
only to volatile anonymous modules and therefore the resulting module
is not installed in the sensor. The rule \Rsend{} (\Rreceive{})
handles the interaction with the network by putting (getting) messages
in (from) the outgoing (incoming) queue. Notice that receiving a 
message is non-blocking (\RreceiveN).
The rules \RcallS{}, \RcallM{} and \Rnocall{} handle calls to
functions in modules. \RcallS{} selects the function in the
sensor's module, gets its code and replaces the parameters with the
arguments passing the sensor's module $M$ as the first argument
in variable \selfk. \RcallM{} is similar to \RcallS{} but uses 
module $M'$ instead of the sensor's module $M$.
Rule \Rnocall{} handles the case of a call to a
function that is not yet installed. The call is deferred to the end of
the run-queue. The idea is that the module containing the function
may not have arrived at the sensor to be installed and so we postpone
the execution of the function.

When a value of $t$ is reached such that it implies the triggering of
a call, the rules \Rtrigger{} and \Rdiscard{} come into action. Rule
\Rtrigger{} places a timed function call $l(\vec v)$ at the front of
the run-queue. The execution of the call is delegated to rule
\RcallS{}. Note that only calls to functions installed in the
sensor~(in $M$) are allowed. Other calls are deferred to the end of the
run-queue by the rule \Rnocall{}. If the timer has expired, rule
\Rdiscard{} removes the corresponding tuple from $T$.


\begin{figure}
  \begin{gather*}
     \\[\rulespace]
     \tag{\Rnetwork, \Rcongr}
     \frac{
       S \reduces S'
    }
    {
      S \parn S''
      \reduces
      S' \parn S''
    }
    \qquad
    \frac{
      S_1 \congr S_2 
      \qquad
      S_2 \reduces S_3
      \qquad
      S_3 \congr S_4
    }{
      S_1 \reduces S_4
    }
     \\[\rulespace]
     \\[\rulespace]
    \tag{\Rbroadcast}
    \frac{
      \dist(p,p') 
      \qquad (I'',O'') = \route(m, I', O')
    }
    {
      \tagsensor {P,R} 
      {M, T} I {m::O} p t S \parn
      \sensor{P',R'}{M', T'} {I'} {O'} {p'}{t'}  
      \reduces
      \tagsensor {P,R}  
      {M, T} I {m::O} p t 
      {S \parn \sensor{P', R'}
        {M', T'} {I''} {O''} {p'}{t'}}
    }
    \\[\rulespace]
     \\[\rulespace]
    \tag{\Rrelease}
    {
      \tagsensor {P,R} {M, T} I {m::O} p t S
      \reduces
      \sensor {P,R} {M, T} I O p t \parn S
    }
\end{gather*}
\caption{Reduction semantics for sensor networks.}
\label{fig:reduction-net}
\end{figure}


Network level reduction proceeds concurrently with in-sensor
processing. It handles the distribution of messages placed by the
sensors in their outgoing queues. 
A message broadcast starts with the creation of an empty membrane for
the broadcasting sensor (rule \SinitSend{} from the structural
congruence). Then, each time a new sensor is added to the membrane of
a broadcasting sensor (rule \Rbroadcast), a function
$\underline\route$ decides where the message in the $O$ queue of the
broadcasting sensor should be copied into the new sensor. The function
can be thought off as implementing the routing protocol for the sensor
network.
The message broadcast ends with the destruction of the membrane, the
captive sensors becoming again free to engage in communication (rule
\Rrelease).



\section{The Type System}
\label{sec:types}

In this section we present a simple type system for Callas, discuss
run-time errors, and prove a type safety result guaranteeing that a
well-typed sensor network does not get ``stuck'' while computing.

\paragraph{Type checking.}
\label{sec:type-checking}
The syntax for types is depicted in Figure~\ref{fig:syntax-types}.
Types $\tau$ are built from the built-in type~$\beta$, the types 
for functions $\arrowTyped$, where~$\vec \tau$ is
the type for parameters of the function and $\tau$ is its return
type, the types for the sensor code module $\typesmodd$ that is a record
type gathering type information for each function of the code module,
the types for anonymous code modules $\typeamodd$, recursive types,
and type variables. The need for distinct code module types comes from
the fact that we need to distinguish from installing code in the sensor module 
or in an anonymous module.
The $\mu$ operator is a binder, giving rise, in the standard way, to
notions of bound and free variables and alpha-equivalence. We do not
distinguish between alpha-convertible types.  Furthermore, we take an
equi-recursive view of
types~\cite{pierce:types-programming-languages}, not distinguishing
between a type $\recTyped$ and its unfolding $\tau\subs\recTyped X$.%

%
%
%


\begin{figure}
\begin{align*}
  & \tau \grmeq & & \text{\emph{Types}}
  \\
  & \qquad \; \beta & & \text{built-in type}
  \\
  & \quad \grmor \arrowTyped & & \text{recursive function type}
  \\
  & \quad \grmor   \typesmodd & & \text{sensor code type}
  \\
  & \quad \grmor   \typeamodd & & \text{anonymous code type}
  \\
  & \quad \grmor   \recTyped & & \text{recursive type}
  \\
  & \quad \grmor   \alpha & & \text{type variable}
\end{align*}
\caption{The syntax of types.}
\label{fig:syntax-types}
\end{figure}


\begin{definition}
  \label{def:plus}
  The $+$ operator is defined (overloaded) for modules, code types, and
  type environment as follows:
  \begin{itemize}
  \item $\obj{l_i = (\vec x_i) P_i}_{i \in I} + \obj{l'_j = (\vec x_j) P'_j}_{j \in J} =
    \obj{\l_i = (\vec x_i) P_i, l'_j = (\vec x_j) P'_j}_{i \in (I \setminus J), j \in J}$
  \item $\obj{l_i \colon \tau_i}_{i \in I} + \obj{l'_j \colon \tau'_j}_{j \in J} =
    \obj{\l_i \colon \tau_i, l'_j \colon \tau'_j}_{i \in (I \setminus J), j \in J}$
  \item $\Gamma_1 + \Gamma_2 = (\Gamma_1 \setminus \Gamma_2) \cup
    \Gamma_2$.
  \end{itemize}
\end{definition}


The typing rules for values, processes, sensors and queues are
presented in Figures~\ref{fig:type-system-values}
to~\ref{fig:type-system-queues}.
Type judgments for values are of the form $\seqSetsd \type v \colon
\tau$, where $\tau_S$ and $\tau_M$ are code module types
representing the types for the built-in functions of the sensor ($\tau_S$) and
for functions installed in the sensor memory $\tau_M$, and $\Gamma$ is a 
typing environment mapping variables to types.
The rules are straightforward, but notice that rule \TVsensor{} assigns the 
sensor code type $\tau_M$ to $\branek$ value.

The judgments for processes are the same as for values.
Rule~\TsCall{} ensures that no user-defined function is executed as a
system call and that a system call always belongs to a predefined
type $\tau_s$ ($\tau_S \type l \colon \functionTyped$).
Broadcasting a call (Rule~\Tbcast{}) is only possible if the call can
be made locally ($\seqSetsd \type \invk \branek l {\vec v} \colon \obj{}$)
and for functions that return the empty module, since it is an
asynchronous remote call and no value is going to be returned
(cf.\@ the return value of a system or a local call, which is synchronous).
Notice that the type system does not distinguish between 
local and remote functions, however such refinement may be 
interesting and can easily be added.
Installing code in the sensor's code module~(Rule~\TinstS) 
implies that the module is entirely replaced and that its type is preserved.
On the other hand, installation over an anonymous 
module~(Rule~\TinstM) is more flexible and only requires that
functions common to both code modules should agree
on their type (\textit{vide} the definition of~$+$ operation).
When calling a local function the type of the first parameter
($\tau_1$) corresponds to the type of module containing the function 
being called (\textit{vide} operation semantic Rules~\RcallS{} and~\RcallM{}
in Figure~\ref{fig:reduction}).
The rules for $\letk$ and $\receivek$ are straightforward.
Finally, firing an event (Rule~\Tfire{}) amounts to calling a user-defined
function locally.

Typing judgments for sensor networks are of the form $\tau_S;
\tau_M \type S$. We only comment the rule for typing a sensor
(Rule~\TSsensor{}), in particular, that the type of each function 
in the sensor's code module~($M$) must agree with predefined 
sensor's type interface~($\tau_M$), apart from the self parameter.

Typing the run-queue (Rule~\TrunQueue,
Figure~\ref{fig:type-system-queues}), the incoming and outgoing
queues, and the event table is equivalent to typing each element of
the structure individually (Rules~\TcommQueue{} and~\TeventQueue{}).
Notice that each element of the incoming (outgoing) queue is typable if it
can be called as a local sensor function. The same holds for timed calls ($\msgd$).


\begin{figure}
  \begin{gather*}
    \seqSetsd \type b \colon \beta
    \qquad
    \seqSetsd \disj x \colon \tau \type x \colon \tau
    \qquad
    \seqSetsd \type \branek \colon \tau_M
    \tag{\Tbin, \TVvar, \TVsensor}
    \\[\rulespace]
    \frac{
      j \in I
    }
    {
      \senObj{l_i \colon \functionType {\vec \tau_i} {\tau_i}}_{i \in I} 
           \type l_j \colon \functionType {\vec \tau_j} {\tau_j}
    }
    \qquad
    \frac{
      \forall i. \seqSetsd \type v_i \colon \tau_i
    }
    {
      \seqSetsd \type \vec v \colon \vec \tau
    }
    \tag{\TVlabel, \TVseq}
    \\[\rulespace]
   \frac{
     \forall i \in I.\seqSetsd
     \disj s_i \colon \tau_{M'} \disj \vec x_i \colon \vec \tau_i \type P_i \colon \tau_i
     \qquad 
     \tau_{M'} = \recType \alpha {\obj{l_i \colon 
         \functionType{\alpha \vec{\tau_i}} \tau_i}_{i \in I}}
   }
   {
     \seqSetsd \type \obj{l_i = {\abstr {s_i,\vec x_i}{P_i}}}_{i \in I} \colon \tau_{M'}
   }
   \tag{\Tcode}
 \end{gather*}
 \center where $\senObj{l_i \colon \functionType {\vec \tau_i} {\tau_i}}_{i
   \in I} $ means either a sensor or an anonymous code type.
\caption{Typing rules 
  for values.}
\label{fig:type-system-values}
\end{figure}



\begin{figure}
  \begin{gather*}
    \tag{\TsCall,\Tbcast}
    \frac{
      \tau_S \type l \colon \functionTyped
      \qquad
      \seqSetsd \type \vec v \colon \vec \tau
    }
    {
      \seqSetsd \type \systemd \colon \tau
    }
    \qquad
    \frac{
      \seqSetsd \type \invk \branek l {\vec v} \colon \obj{}
    }
    {
      \seqSetsd \type \sendd \colon \obj{}
    }
    \\[\rulespace]    
    \tag{\TinstS, \TinstM}
    \frac{
      \begin{array}{c}
        \seqSetsd \type v_1 \colon 
           \recType \alpha {\anyObj {l_i \colon \alpha \vec \tau_i \rightarrow \tau_i}_{i \in I}}\\
        \seqSetsd \type v_2 \colon 
           \recType \alpha {\obj {l_i \colon \alpha \vec \tau_i \rightarrow \tau_i}_{i \in I}}
      \end{array}
    }
    {
      \seqSetsd \type \install {v_1} {v_2} \colon \obj{}
    }
    \qquad
    \frac{
     \begin{array}{c}
       \seqSetsd \type v_1 \colon \tau_1
       \qquad
       \tau_1 = \obj{l_i \colon \vec \tau_i \rightarrow \tau_i}_{i \in I}\\
       \seqSetsd \type v_2 \colon \tau_2
       \qquad
       \tau_2 = \obj{l_j \colon \vec \tau_j \rightarrow \tau_j}_{j \in J}
     \end{array}
   }
    {
      \seqSetsd \type \install {v_1} {v_2} \colon \tau_1 + \tau_2
    }
    \\[\rulespace]
   \tag{\Tcall,\Tlet}
    \frac{
      \seqSetsd \type v_1 \colon \tau_1
      \quad
      \tau_1 \type l \colon \functionType {\tau_1 \vec \tau}{\tau_2}
      \quad 
      \seqSetsd \type \vec v_2 \colon \vec \tau
    }
    {
      \seqSetsd \type \invk {v_1} {l} {\vec v_2} \colon \tau_2
    }
    \qquad
    \frac{
      \seqSetsd \type P_1 \colon \tau_1
      \quad
      \seqSetsd \disj x \colon \tau_1 \type P_2 \colon \tau_2
    }
    {
      \seqSetsd \type \Let x {P_1} {P_2} \colon \tau_2
    }
    \\[\rulespace]
    \tag{\Treceive,\Tfire}
    \seqSetsd \type \receivedd \colon \obj{}
    \qquad
    \frac{
      \seqSetsd \type \invk \branek l {\vec v} \colon \obj{}
      \qquad
      \seqSetsd \type v_1 v_2 \colon \beta \beta
    }
    {
      \seqSetsd \type \fire {\msgd} {v_1} {v_2} \colon \obj{}
    }
 \end{gather*}
\caption{Typing rules for processes.}
\label{fig:type-system-processes}
\end{figure}



\begin{figure}
  \begin{gather*}
   \tau_S;\tau_M \type \inaction
   \qquad
   \frac{
     \begin{array}{c}
       \tau_S;\tau_M;\emptyset \type P \colon \_
       \quad
       \tau_S;\tau_M \type R 
       \quad
       \tau_S;\tau_M;\emptyset \type M \colon 
       \recType \alpha 
          {\obj{l_i \colon \functionType {\alpha \vec \tau_i}
              {\tau_i}}_{i \in I}}
       \\
       \forall i \in I.\tau_M \type l_i \colon 
       \functionType {\tau_M \vec \tau_i} {\tau_i}
       \qquad
       \tau_S;\tau_M; \emptyset \type t p \colon \beta \beta
       \\
       \tau_S;\tau_M \type T 
       \qquad 
       \tau_S;\tau_M \type I 
       \qquad 
       \tau_S;\tau_M \type O 
     \end{array}
   }
   {
     \tau_S;\tau_M \type \sensord
   }
   \tag{\TSinaction, \TSsensor}
   \\[\rulespace]
   \frac{
     \tau_S;\tau_M \type \sensord
     \qquad
     \tau_S;\tau_M \type S
   }
   {
     \tau_S;\tau_M \type \tagsensord 
   }
   \qquad
   \frac{
     \tau_S;\tau_M \type S_1 
     \qquad
     \tau_S;\tau_M \type S_2
   }
   {
     \tau_S;\tau_M \type S_1 \parn S_2
   }
   \tag{\TSbSensor, \Tparallel}
 \end{gather*}
\caption{Typing rules for sensor networks.}
\label{fig:type-system-sensors}
\end{figure}


\begin{figure}
  \begin{gather*}
    \frac{
      \tau_s;\tau_M;\emptyset \type P \colon \_
      \qquad
      \tau_s;\tau_M \type R
    }
    {
      \tau_S;\tau_M \type P::R
    }
    \qquad \qquad
    \frac{
      \tau_S;\tau_M;\emptyset \type \invk \branek l {\vec v} \colon \_
      \qquad
      \tau_S;\tau_M \type I
    }
    {
      \tau_S;\tau_M \type \langle \msgd \rangle :: I
    }
    \tag{\TrunQueue, \TcommQueue}
    \\[\rulespace]
    \frac{
      \tau_S;\tau_M;\emptyset \type \invk \branek l {\vec v} \colon \_
      \qquad
      \tau_S;\tau_M;\emptyset \type v_1 v_2 v_3 \colon \beta_1 \beta_2 \beta_3
      \qquad
      \tau_S;\tau_M \type T
    }
    {
      \tau_S;\tau_M \type T \uplus (\msgd, v_1, v_2, v_3) 
    }
    \tag{\TeventQueue}
  \end{gather*}
  \caption{Typing rules for queues of messages, processes, and events.}
  \label{fig:type-system-queues}
\end{figure}

 
The proofs for our main results (Theorem~\ref{teo:subject-reduction},
Theorem~\ref{teo:type-safety}, and Corollary~\ref{cor:type-safety})
are based on the following auxiliary results.
We call \textit{context process}, denoted $\context{P}$, the
processes resulting from filling its context hole.
Informally, Lemma~\ref{lemma:context-process} states that if a context
process is well typed, then the same also holds for the process that
fills its hole, although not necessarily with an identical type.
Lemma~\ref{lemma:process-context} states that the typability of a
context process holds and its type is preserved if we fill the context's
hole with processes of the same type.
Lemma~\ref{lemma:module-substitution} handles module's substitution.
Lemmas~\ref{lemma:substitution-lemma} and~\ref{lemma:congruence} are
discussed below.

\begin{lemma}
  \label{lemma:context-process}
  If $\tau_S; \tau_M;\Gamma \type \context P \is \tau$, then $\tau_S;
  \tau_M;\Gamma' \type P \is \tau'$.
\end{lemma}

\begin{proof}
  The proof proceeds by induction on the contexts' structure and
  both cases are straightforward. 
\end{proof}

\begin{lemma}
  \label{lemma:process-context}
  If $\tau_S; \tau_M;\Gamma \type \context P \is \tau$, $\tau_S;
  \tau_M; \Gamma' \type P \is \tau'$, and $\tau_S; \tau_M;\Gamma' \type P' \is
  \tau'$, then $\tau_S; \tau_M;\Gamma \type \context {P'} \is \tau$.
\end{lemma}

\begin{proof}
  We proceed by induction on the contexts' structure analysing each 
  definition case. Both cases follow easily.
\end{proof}

\begin{lemma}
  \label{lemma:module-substitution}
  If $\tau_S; \tau_M;\Gamma \type M_1 \is \tau_1$ and $\tau_S; \tau_M;\Gamma'
  \type M_2 \is \tau_2$, 
  then $\tau_S; \tau_M; \Gamma + \Gamma' \type M_1 \methJoin 
  M_2 \is \tau_1 \methJoin \tau_2$.
\end{lemma}

\begin{proof}
  Directly from the definition of $\methJoin$ and using Rule~\Tcode.
\end{proof}

The Substitution Lemma is used in the proof of the Subject Reduction
Theorem, to show the cases that involve the replacement of formal by
actual parameters, specifically for function call and for the let
construct.
The proof is standard, so we omit it, but the interested reader may
find similar proofs in the literature, for instance
in~\cite[Section 6.3]{sangiorgi.walker:theory-mobile}.

\begin{lemma}[Substitution Lemma]
  \label{lemma:substitution-lemma}
  If $\tau_S; \tau_M;\Gamma \type v \is \tau'$ and $\tau_S; \tau_M;\Gamma
  \disj x \is \tau' \type P \is \tau$, then $\tau_S; \tau_M;\Gamma \type
  P \subs v x \is \tau$.
\end{lemma}

The following results state type invariance during reduction.

\begin{lemma}[Congruence Lemma]
  \label{lemma:congruence}
  If $\tau_S; \tau_M \type S$ and $S \congr S'$, then $\tau_S;
  \tau_M \type S'$.
\end{lemma}

\begin{proof}
  We proceed by induction on the derivation tree for $S \congr S'$.
  The proof is straightforward.
\end{proof}

\begin{theorem}[Subject Reduction]
  \label{teo:subject-reduction}
  If $\tau_S; \tau_M \type S$ and $S \reduces S'$, then $\tau_S;
  \tau_M \type S'$.
\end{theorem}

\begin{proof}
  By induction on the derivation tree for $S \reduces S'$. 
  In each case, we proceed by case analysis on the last typing rule of
  the inference tree for $\tau_S; \tau_M \type S$.
\end{proof}

\paragraph{Type safety.}
\label{sec:type-safety}
Our claim is that well-typed sensor networks are free from run-time errors.
The unary relation $S \err$, defined as the least relation on
networks closed under the rules in Figure
\ref{fig:runtime-errors}, identifies processes that would get ``stuck''
during computation (reduction).
We write $S \nerr$ for $\lnot (S \err)$.

Our Sensor Networks may exhibit two kinds of failures upon
computing: when calling a function or when installing a module.
In the former, the call may result in a run-time error when the target
of the call is neither \lstinline{sensor}, nor an anonymous module (Rule
\Ecall);
or when the function name is unknown or there is a mismatch
between the number of arguments ($v_1 \dots v_n$) and the number
parameters ($x_1 \dots x_m$) (Rule \Efunction).
In the latter, an error may occur if we are installing some value
that is not a module (Rule \Einstall).

As an example, recall the \lstinline{gather} function from the
streaming data example that we sketched below.
\begin{lstlisting}[frame=none]
  { gather = (self,x,y) ...}
\end{lstlisting}
The process
\begin{lstlisting}[frame=none]
  let t = extern getTime() in send gather(t)
\end{lstlisting}
exhibits a run-time error, since function
\lstinline|gather| is being called with two arguments instead of three.
In fact, the above network may reduce using Rules~\Rsystem{} and~\Rlet, but
then we cannot apply Rule \Rfunction, since the substitution is not
defined.
Run-time error Rule \Efunction{} captures this kind of failure.


\begin{figure}
  \begin{align*}
  \sensor {\context{\invk v l {\vec v}}, R} {M,T} I O p t  & \err & &
  \text{if } v \text{ is not \lstinline{sensor}, nor } M'
  \tag{\Ecall}\\[0.07cm]
  \sensor {\context{\invk {M'} l {v_1 \dots v_n}}, R} {M,T} I O p t & \err & &
  \text{if } l \not \in \dom(M') \text{ or }
  \\
  & & &  (M'(l) = \abstr {x_1 \dots x_m} {P} \text{ and }\\
  & & & n \neq m)
    \tag{\Efunction}\\[0.07cm]
    \sensor {\context {\install v v}, R} {M, T} I O p t & \err & &
    \text{if } v \text{ is not \lstinline{sensor}, nor } M'
    \tag{\Einstall}\\[-1cm]
  \end{align*}
\begin{align*}
  \frac{
    S \err
  }
  {
    S \parn S' \err
  }
  \qquad \qquad
  \frac{
    S \congr S' \quad
    S \err
  }
  {
    S' \err
  }
  \tag{\Epar, \Estr}
\end{align*}
\caption{Run-time errors for sensors.}
\label{fig:runtime-errors}
\end{figure}


The Type Safety result states that well-typed networks do not incur in
run-time errors.

\begin{theorem}[\bf Type Safety]
  \label{teo:type-safety}
  If $\tau_S;\tau_M \type S$, then $S \nerr$.
\end{theorem}

\begin{proof}
  We prove the contra-positive result, namely $S \err$ implies that
  $\tau_S;\tau_M \not \type S$, proceeding by induction on the
  definition of $S \err$ relation.
\end{proof}

Finally, a well-typed network is free of flaws, at any time during
reduction.

\begin{corollary}[\bf Absence of Runtime Errors]
  \label{cor:type-safety}
  If $\tau_S;\tau_M \type S$ and $S \reducesn S'$, then $S' \nerr$.
\end{corollary}

\begin{proof}
  By hypothesis $\tau_S;\tau_M \type S$, then, since types are
  preserved during reduction (Theorem \ref{teo:subject-reduction}), by
  induction on the length of $\reducesn$ we obtain $\tau_S;\tau_M
  \type S'$.
  Using the Type Safety theorem (Theorem \ref{teo:type-safety}) we
  conclude that $S' \nerr$.
\end{proof}



\section{Related work}
\label{sec:related}

The majority of available programming tools for
sensor networks are based on rather low-level programming languages,
most notably the module-based idiom nesC~\cite{nesc:gay:levis:etal},
which promotes a system level programming style on top of a
small-scale operating system such as TinyOS~\cite{tinyos}. Other
examples include C and Prothothreads for Contiki~\cite{contiki} and at
the extreme Pushpin~\cite{pushpin}.

Moving away from the hardware and system level programming we have
virtual machines like Mat\'{e}~\cite{mate:levis:culler:02} and its
associated core language TinyScript that provide programmers with a
suitable abstraction layer for the hardware.
Middleware platforms such as Deluge~\cite{deluge:hui:culler:04} and
Agilla~\cite{agilla:fok:roman:lu:05} enable higher level control of
sensor networks for critical operations such as massive code
deployment.

True high-level programming languages such as Regiment~\cite{regiment},
Cougar~\cite{cougar}, and TinyDB~\cite{tinydb} abstract away from the
physical network by viewing sensor networks as time varying data
streams or as data repositories. 
Regiment, for instance, adopts a data-centric view of sensor networks
and provides the programmer with abstractions to manipulate data
streams and to manage network regions.
Although Regiment is a strongly typed language --- an essential
characteristic to enable the scalable development of applications
--- its construction is not based on a formal calculus and it is not
clear that the semantics is amenable to proving correctness results
for the system and applications.

In fact, the state-of-the-art in the design of sensor network
programming languages~\cite{bookchapter} follows, invariably, a
\emph{top-down} approach, in which system engineers start by
identifying useful patterns and abstractions based on case studies of
applications and then attempt to provide the programmer with language
constructs and system features that reflect these patterns.  These
building blocks must then be compiled into nesC/TinyOS code or some
other API that interacts with the low-level operating system.
The problem with such approaches is that the semantic gap between
the original language specification and the actual implementation
inevitably precludes a thorough analysis of the correctness of
the envisioned sensor networking application.

Seeking a fundamentally sound path towards the development of
programming languages for sensor networks, we propose
a somewhat disruptive \emph{bottom-up} approach.
Inspired by process calculi
theory~\cite{async-pi:honda:tokoro:91,pi:milner:parrow:walker:92}, our
basic idea is to start by constructing a fundamental programming
model, which (a) captures the specific computing and communication
aspects of sensor networks and (b) enables us to reason about their
fundamental operations.
This approach is justified by the fact that most high-level languages,
even those that fully abstract from the networking aspects and view
sensor networks as time varying data streams or data repositories,
ultimately map their high-level constructs into a lower level
communication-centric language and run-time system.

Previous work on process calculi for wireless systems is scarce.
Prasad~\cite{broadcast:prasad:91} established the first process
calculus approach to modeling broadcast based systems.
Later work by Ostrovsk\'y, Prasad, and
Taha~\cite{broadcast-high-order:ostrovsky:prasad:taha:02} established
the basis for a higher-order calculus for broadcasting systems.
More recently, Mezzetti and
Sangiorgi~\cite{wireless:mezzetti:sangiorgi:06} discuss the use of
process calculi to model wireless systems, again focusing on the
details of the lower layers of the protocol stack
(\emph{e.g.\@}~collision avoidance) and establishing an operational
semantics for the networks.

\section{Conclusions}
\label{sec:conclusions}

\paragraph{Discussion.}
\label{sec:discussion}

Programming languages based on type safe specifications are
fundamental for applications where development and debugging can be
complex. Sensor networks are one such case. Difficulties in physically
accessing deployed sensors, resource limitations of the devices, and
dynamic ad-hoc routing protocols, all conspire to make the programming
and debugging of these infrastructures a difficult task.

In this paper we present a strongly typed calculus for programming
sensor networks. Sensor network applications are built by plugging
together components called \emph{modules}. Dynamic reprogramming 
is supported by making modules first class entities that
can be exchanged between sensors and by allowing modules to be
installed locally upon reception on a sensor. A type system provides a
static verification tool, which allows for premature detection of
protocol errors in the usage of modules. This feature is of utmost
importance when programming large-scale applications for sensor
networks, since it eliminates many errors that would have to be
corrected online, at run-time. We prove two fundamental
properties of the operational semantics and of the type system,
namely, \emph{subject reduction} and \emph{type safety}. Together,
these results establish the calculus as a sound framework for
developing programming languages for sensor networks.

\paragraph{Future work.}
As part of our ongoing work, we are pursuing two different lines of
research. First, we are exploring the theoretical properties of the
calculus. By applying techniques from process calculi theory we hope
to be able to prove fundamental properties of sensor networking
applications and protocols (\eg protocol correctness). Second, we
designed a core programming language based on the calculus and
implemented the corresponding compiler and virtual machine. We expect
to prove the correctness of the virtual machine relative to the base
calculus.  This will provide an unequivocal link between the semantics
of the calculus (and core language) and the semantics of higher-level
programming languages that we implement on top of it.


~\vspace{-0.5cm}
\paragraph{Acknowledgments.}
The authors are partially supported by project CALLAS of the
Funda\c{c}\~ao para a Ci\^encia e Tecnologia (contract
PTDC/EIA/71462/2006).\\[-0.8cm]


\begin{thebibliography}{10}
\providecommand{\bibitemstart}[1]{\bibitem{#1}}
\providecommand{\bibitemend}{}
\providecommand{\bibliographystart}{}
\providecommand{\bibliographyend}{}
\providecommand{\url}[1]{\texttt{#1}}
\providecommand{\urlprefix}{Available at }
\providecommand{\bibinfo}[2]{#2}
\bibliographystart

\bibitemstart{survey:akyildiz:etal:02}
\bibinfo{author}{I.~Akyildiz}, \bibinfo{author}{W.~Su},
  \bibinfo{author}{Y.~Sankarasubramaniam} \& \bibinfo{author}{E.~Cayirci}
  (\bibinfo{year}{2002}): \emph{\bibinfo{title}{{A Survey on Sensor
  Networks}}}.
\newblock {\sl \bibinfo{journal}{IEEE Communications Magazine}}
  \bibinfo{volume}{40}(\bibinfo{number}{8}), pp. \bibinfo{pages}{102--114}.
\bibitemend

\bibitemstart{contiki}
\bibinfo{author}{A.~Dunkels}, \bibinfo{author}{B.~Gr\"{o}nvall} \&
  \bibinfo{author}{T.~Voigt} (\bibinfo{year}{2004}):
  \emph{\bibinfo{title}{{Contiki - a Lightweight and Flexible Operating System
  for Tiny Networked Sensors}}}.
\newblock In: {\sl \bibinfo{booktitle}{EmNets'04 Workshop}}.
\bibitemend

\bibitemstart{agilla:fok:roman:lu:05}
\bibinfo{author}{C.-L. Fok}, \bibinfo{author}{G.-C. Roman} \&
  \bibinfo{author}{C.~Lu} (\bibinfo{year}{2005}): \emph{\bibinfo{title}{{Rapid
  Development and Flexible Deployment of Adaptive Wireless Sensor Network
  Applications}}}.
\newblock In: {\sl \bibinfo{booktitle}{{ICDCS'05}}}. \bibinfo{publisher}{IEEE
  Press}, pp. \bibinfo{pages}{653--662}.
\bibitemend

\bibitemstart{cougar}
\bibinfo{author}{W.~F. Fung}, \bibinfo{author}{D.~Sun} \&
  \bibinfo{author}{J.~Gehrke} (\bibinfo{year}{2002}):
  \emph{\bibinfo{title}{{COUGAR}: The Network is the Database}}.
\newblock In: {\sl \bibinfo{booktitle}{SIGMOD'02}}. \bibinfo{publisher}{ACM
  Press}.
\bibitemend

\bibitemstart{nesc:gay:levis:etal}
\bibinfo{author}{D.~Gay}, \bibinfo{author}{P.~Levis}, \bibinfo{author}{R.~von
  Behren}, \bibinfo{author}{M.~Welsh}, \bibinfo{author}{E.~Brewer} \&
  \bibinfo{author}{D.~Culler} (\bibinfo{year}{2003}):
  \emph{\bibinfo{title}{{The nesC Language: A Holistic Approach to Network
  Embedded Systems}}}.
\newblock In: {\sl \bibinfo{booktitle}{{PLDI'03}}}. \bibinfo{publisher}{{ACM}
  Press}, pp. \bibinfo{pages}{1--11}.
\bibitemend

\bibitemstart{async-pi:honda:tokoro:91}
\bibinfo{author}{K.~Honda} \& \bibinfo{author}{M.~Tokoro}
  (\bibinfo{year}{1991}): \emph{\bibinfo{title}{{A}n {O}bject {C}alculus for
  {A}synchronous {C}ommunication}}.
\newblock In: {\sl \bibinfo{booktitle}{ECOOP'91}}, number \bibinfo{number}{512}
  in \bibinfo{series}{LNCS}. \bibinfo{publisher}{Springer-Verlag}, pp.
  \bibinfo{pages}{133--147}.
\bibitemend

\bibitemstart{deluge:hui:culler:04}
\bibinfo{author}{J.~W. Hui} \& \bibinfo{author}{D.~Culler}
  (\bibinfo{year}{2004}): \emph{\bibinfo{title}{{The Dynamic Behavior of a Data
  Dissemination Protocol for Network Programming at Scale}}}.
\newblock In: {\sl \bibinfo{booktitle}{{ENSS'04}}}. \bibinfo{publisher}{{ACM}
  Press}, pp. \bibinfo{pages}{81--94}.
\bibitemend

\bibitemstart{mate:levis:culler:02}
\bibinfo{author}{P.~Levis} \& \bibinfo{author}{D.~Culler}
  (\bibinfo{year}{2002}): \emph{\bibinfo{title}{{Mat\'{e}: A Tiny Virtual
  Machine for Sensor Networks}}}.
\newblock In: {\sl \bibinfo{booktitle}{{ASPLOS X}}}. \bibinfo{publisher}{{ACM}
  Press}, pp. \bibinfo{pages}{85--95}.
\bibitemend

\bibitemstart{pushpin}
\bibinfo{author}{J.~Lifton}, \bibinfo{author}{D.~Seetharam},
  \bibinfo{author}{M.~Broxton} \& \bibinfo{author}{J.~Paradiso}
  (\bibinfo{year}{2002}): \emph{\bibinfo{title}{{Pushpin Computing System
  Overview: a Platform for Distributed}}}.
\newblock In: {\sl \bibinfo{booktitle}{Pervasive'02}}.
  \bibinfo{publisher}{Springer-Verlag}.
\bibitemend

\bibitemstart{bookchapter}
\bibinfo{author}{L.~Lopes}, \bibinfo{author}{F.~Martins} \&
  \bibinfo{author}{J.~Barros} (\bibinfo{year}{2009}):
  \emph{\bibinfo{title}{{Middleware for Network Eccentric and Mobile
  Applications}}}, chapter~\bibinfo{chapter}{2}, pp. \bibinfo{pages}{25--41}.
\newblock \bibinfo{publisher}{Springer-Verlag}.
\bibitemend

\bibitemstart{sensorcomm07}
\bibinfo{author}{L.~Lopes}, \bibinfo{author}{F.~Martins},
  \bibinfo{author}{M.~S. Silva} \& \bibinfo{author}{João Barros}
  (\bibinfo{year}{2007}): \emph{\bibinfo{title}{{A Process Calculus Approach to
  Sensor Network Programming}}}.
\newblock In: {\sl \bibinfo{booktitle}{SENSORCOMM '07}}.
  \bibinfo{publisher}{IEEE Computer Society}, pp. \bibinfo{pages}{451--456}.
\bibitemend

\bibitemstart{tinydb}
\bibinfo{author}{S.~Madden}, \bibinfo{author}{M.~J. Franklin},
  \bibinfo{author}{J.~M. Hellerstein} \& \bibinfo{author}{W.~Hong}
  (\bibinfo{year}{2005}): \emph{\bibinfo{title}{{TinyDB: An Acquisitional Query
  Processing System for Sensor Networks}}}.
\newblock {\sl \bibinfo{journal}{ACM Transactions on Database Systems}} .
\bibitemend

\bibitemstart{wireless:mezzetti:sangiorgi:06}
\bibinfo{author}{N.~Mezzetti} \& \bibinfo{author}{D.~Sangiorgi}
  (\bibinfo{year}{2006}): \emph{\bibinfo{title}{{Towards a Calculus for
  Wireless Systems}}}.
\newblock In: {\sl \bibinfo{booktitle}{{MFPS}'06}}, {\sl
  \bibinfo{series}{ENTCS}} \bibinfo{volume}{158}. \bibinfo{publisher}{Elsevier
  Science}, pp. \bibinfo{pages}{331--354}.
\bibitemend

\bibitemstart{ccs:milner:80}
\bibinfo{author}{R.~Milner} (\bibinfo{year}{1980}): \emph{\bibinfo{title}{{A
  Calculus of Communicating Systems}}}.
\newblock Number~\bibinfo{number}{92} in \bibinfo{series}{LNCS}.
  \bibinfo{publisher}{Springer-Verlag}.
\bibitemend

\bibitemstart{pi:milner:parrow:walker:92}
\bibinfo{author}{R.~Milner}, \bibinfo{author}{J.~Parrow} \&
  \bibinfo{author}{D.~Walker} (\bibinfo{year}{1992}): \emph{\bibinfo{title}{{A}
  {C}alculus of {M}obile {P}rocesses, ({P}arts {I} and {II})}}.
\newblock {\sl \bibinfo{journal}{Information and Computation}}
  \bibinfo{volume}{100}, pp. \bibinfo{pages}{1--77}.
\bibitemend

\bibitemstart{regiment2}
\bibinfo{author}{R.~Newton}, \bibinfo{author}{Arvind} \&
  \bibinfo{author}{M.~Welsh} (\bibinfo{year}{2005}):
  \emph{\bibinfo{title}{{Building up to Macroprogramming: An Intermediate
  Language for Sensor Networks.}}}
\newblock In: {\sl \bibinfo{booktitle}{IPSN'05}}. pp. \bibinfo{pages}{37--44}.
\bibitemend

\bibitemstart{regiment}
\bibinfo{author}{R.~Newton} \& \bibinfo{author}{M.~Welsh}
  (\bibinfo{year}{2004}): \emph{\bibinfo{title}{{Region Streams: Functional
  Macroprogramming for Sensor Networks}}}.
\newblock In: {\sl \bibinfo{booktitle}{{DMSN'04} Workshop}}.
\bibitemend

\bibitemstart{broadcast-high-order:ostrovsky:prasad:taha:02}
\bibinfo{author}{K.~Ostrovsk\'y}, \bibinfo{author}{K.~V.~S. Prasad} \&
  \bibinfo{author}{W.~Taha} (\bibinfo{year}{2002}):
  \emph{\bibinfo{title}{{Towards a Primitive Higher Order Calculus of
  Broadcasting Systems}}}.
\newblock In: {\sl \bibinfo{booktitle}{{PPDP'02}}}. \bibinfo{publisher}{{ACM}
  Press}, pp. \bibinfo{pages}{2--13}.
\bibitemend

\bibitemstart{pierce:types-programming-languages}
\bibinfo{author}{B.~C. Pierce} (\bibinfo{year}{2002}):
  \emph{\bibinfo{title}{Types and Programming Languages}}.
\newblock \bibinfo{publisher}{MIT Press}.
\bibitemend

\bibitemstart{broadcast:prasad:91}
\bibinfo{author}{K.~V.~S. Prasad} (\bibinfo{year}{1991}):
  \emph{\bibinfo{title}{{A Calculus of Broadcasting Systems}}}.
\newblock In: {\sl \bibinfo{booktitle}{{TAPSOFT}'91}}, number
  \bibinfo{number}{493} in \bibinfo{series}{LNCS}.
  \bibinfo{publisher}{Springer-Verlag}, pp. \bibinfo{pages}{338--358}.
\bibitemend

\bibitemstart{sangiorgi.walker:theory-mobile}
\bibinfo{author}{D.~Sangiorgi} \& \bibinfo{author}{D.~Walker}
  (\bibinfo{year}{2001}): \emph{\bibinfo{title}{The $\pi$-calculus: a Theory of
  Mobile Processes}}.
\newblock \bibinfo{publisher}{Cambridge University Press}.
\bibitemend

\bibitemstart{tinyos}
\bibinfo{author}{TinyOS}.
\newblock \emph{\bibinfo{title}{{The TinyOS Documentation Project}}}.
\newblock \bibinfo{note}{Available at \textsf{http://www.tinyos.org}}.
\bibitemend

\bibliographyend
\end{thebibliography}



\end{document}